\documentclass[reprint,
superscriptaddress,aps,pre,twocolumn,floatfix,showpacs,amsmath,10pt]{revtex4-1}

\usepackage{booktabs}
\usepackage{xcolor}
\usepackage{graphicx}
\usepackage{graphics}
\usepackage{amsmath}
\usepackage{amssymb}
\usepackage{amsthm,amstext}
\usepackage{hyperref}
\usepackage{amsfonts}
\usepackage{lipsum}
\usepackage{MnSymbol}
\usepackage{mathrsfs}
\usepackage{stmaryrd}
\usepackage{latexsym}
\usepackage[applemac]{inputenc}
\usepackage{accents}

\usepackage[bbgreekl]{mathbbol}
\DeclareSymbolFontAlphabet{\mathbbm}{bbold}
\DeclareSymbolFontAlphabet{\mathbb}{AMSb}
\DeclareMathAlphabet\mathbfcal{OMS}{cmsy}{b}{n}

\renewcommand\d\delta
\newcommand\D\Delta

\newcommand\beq{\begin{equation}}
\newcommand\beqn{\begin{eqnarray}}
\newcommand\eeq{\end{equation}}
\newcommand\eeqn{\end{eqnarray}}

\usepackage[normalem]{ulem}

\begin{document}

\title{The Poynting effect}

\author{Giuseppe Zurlo$^a$, James Blackwell$^{b,a}$, Niall Colgan$^b$, Michel Destrade$^a$\\[12pt]
$^a$School of Mathematics, Statistics and Applied Mathematics; $^b$School of Physics;\\
NUI Galway, University Road, Galway, Ireland.}

\date{\today}

\begin{abstract}
An  elastic cylinder subjected to torsion will always elongate.
\end{abstract}

\maketitle


\section{The Poynting effect}

\begin{figure}[htb]
\begin{centering}
\includegraphics[width=\columnwidth]{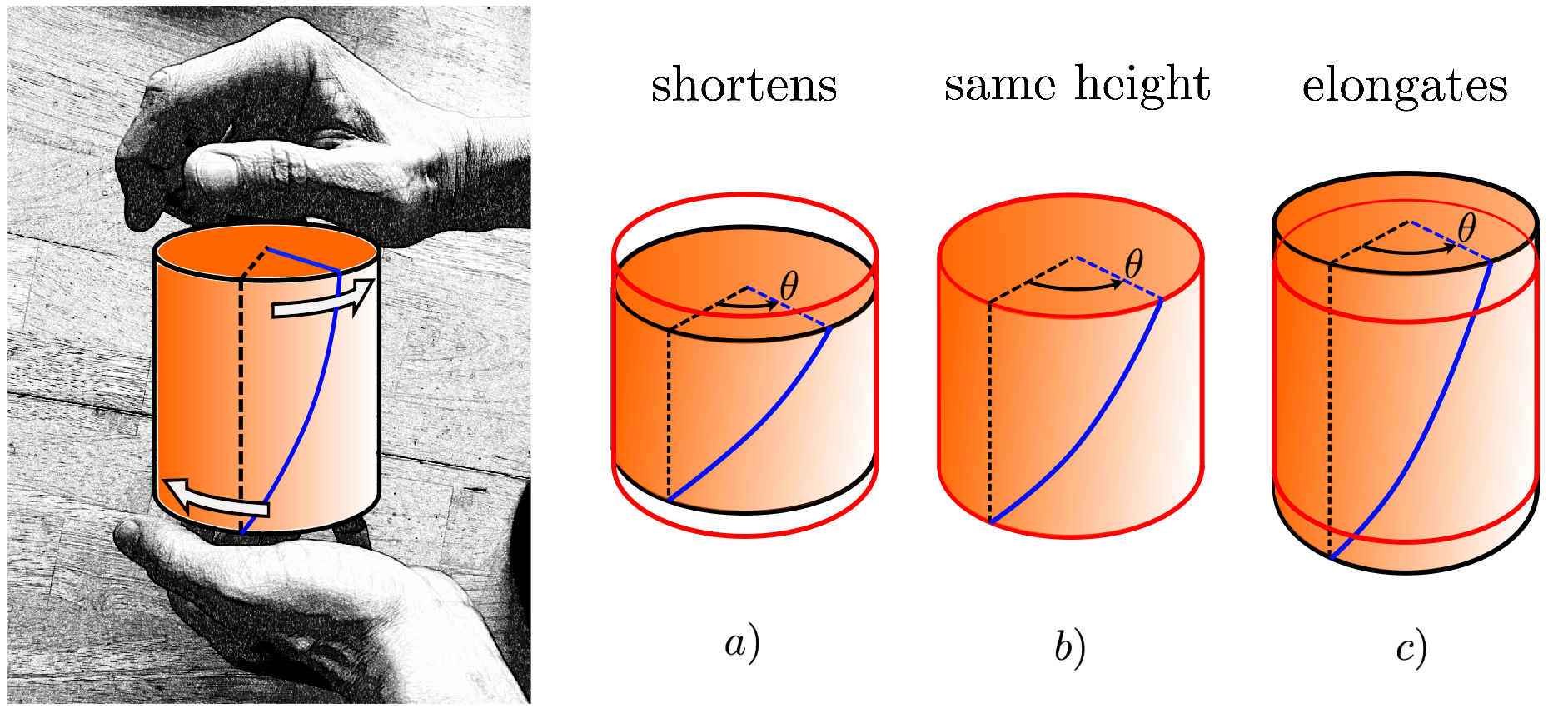}
\caption{\label{cylinders}}
\end{centering}
\end{figure}

Take a solid cylinder and twist it to produce a rotation of angle $\theta$ between the upper and lower faces, see  Fig.\ref{cylinders}. 
Do you expect the cylinder to $a)$ shorten, $b)$ keep its height, or $c)$ elongate? 
The answer to this seemingly innocuous question is not easy to guess: we conducted opinion polls to scientifically-minded populations (science and engineering undergraduates, continuous development students, science festival audiences) and found that their replies were roughly equally divided among the three choices. 

The correct answer is $c)$, which is clearly not intuitive, especially when we think of everyday experimentations, such as twisting an empty aluminium can to crush it. 


\section{A historical perspective}


In the 1911 edition of the Encyclopaedia Britannica \cite{Britannica}, the British physicist John Henry Poynting describes the famous ``Cavendish experiment''  (1797-1798). 
It was conceived (following a concept devised earlier by the geophysicist John Michell) to measure the force of gravity between two masses with a torsion balance: a light bar connecting  two spherical masses is suspended from its middle by a thin and long fiber, and acts as a very sensitive torsion spring as the small spheres are  attracted to larger, much heavier balls. 

Earlier, in 1785,  the French physicist Charles-Augustin de Coulomb had used a torsion balance to assess experimentally what is now known as ``Coulomb's inverse-square law'',  quantifying the electrostatic force between two stationary, electrically charged particles. 
Remarkably, the previous year, Coulomb \cite{Coulomb} had already derived the first mathematical model for the mild torsion of metal wires, by establishing that the torque is \textit{proportional to the torsional angle, the fourth power of the wire diameter and the inverse of the length of the wire}, a law that he verified experimentally.
 
Torsion, as it appears, received a lot of attention during the 18$^\text{th}$ Century. 
Nonetheless, the fact that a wire changes its length when twisted seems to have escaped the attention of all until, in 1909, Poynting \cite{Poynting1909} showed it always elongates. 
Explaining and modelling the so-called \textit{Poynting effect} requires the development of a sound theory of non-linear elasticity, and it then took another four decades for Rivlin \cite{Rivlin47} to propose an analytical solution.


\section{Energy Minimization: A Primer}
\label{Energy Minimization: A Primer}


To prove the existence of the Poynting effect with a minimal technical baggage, we will use the \textit{principle of energy minimization}, which states that when an elastic body is subjected to the action of forces, it deforms so as to minimise a suitable ``energy''. This principle can be easily illustrated with a simple example.

Take a one-dimensional spring, whose state is completely characterised by its length.  Call $x_0$ its stress-free length, and $x$ its length upon the application of a force $F$. 
The British {\it natural philosopher} 
(as applied mathematicians were called back in the 17$^\text{th}$ century) Robert Hooke established that the equilibrium length $x^\star$ of the spring is found by solving 
\beq\label{Hooke}
F = k (x^\star -x_0),
\eeq
where the positive constant $k$ is the spring stiffness. 
The relation above now bears, indeed, the name of ``Hooke's law'' (interestingly, Hooke safeguarded its discovery by publishing it under the anagram {\it ceiiinosssttuv} of the Latin sentence {\it ut tensio, sic vis}, which translates as {\it as the extension, so the force}). 

Remarkably, the equilibrium length $x^\star= x_0 + F/k$ found from \eqref{Hooke} minimises the {\it total potential energy} of the spring, which can be defined as follows. 
First, introduce the spring's {\it elastic energy},
\beq
U = \tfrac{1}{2} k(x -x_0)^2,
\eeq
and the mechanical work done by the force $F$,
\beq
W = F(x-x_0). 
\eeq
Then the  {\it total potential energy} is defined as
\beq
E = U - W. 
\eeq
If the force $F$ is prescribed, we can now show that the equilibrium length $x^\star= x_0 + F/k$ determined above is a minimizer of the energy $E(x)$, in the sense that $E(x^\star)\leq E(x)$ for all $x\neq x^\star$. Indeed, we find the minimum of $E$ from
\beq\label{cond}
\frac{d E}{dx} = k(x-x_0)-F = 0, \qquad 
\frac{d^2 E}{dx^2} =k \geq 0. 
\eeq
confirming that $x^\star$ is a minimiser of $E$, and that the stiffness $k$ is positive. 

%


\section{Poynting effect in torsion}


\begin{figure}[htb]
\begin{centering}
\includegraphics[width=7.5cm]{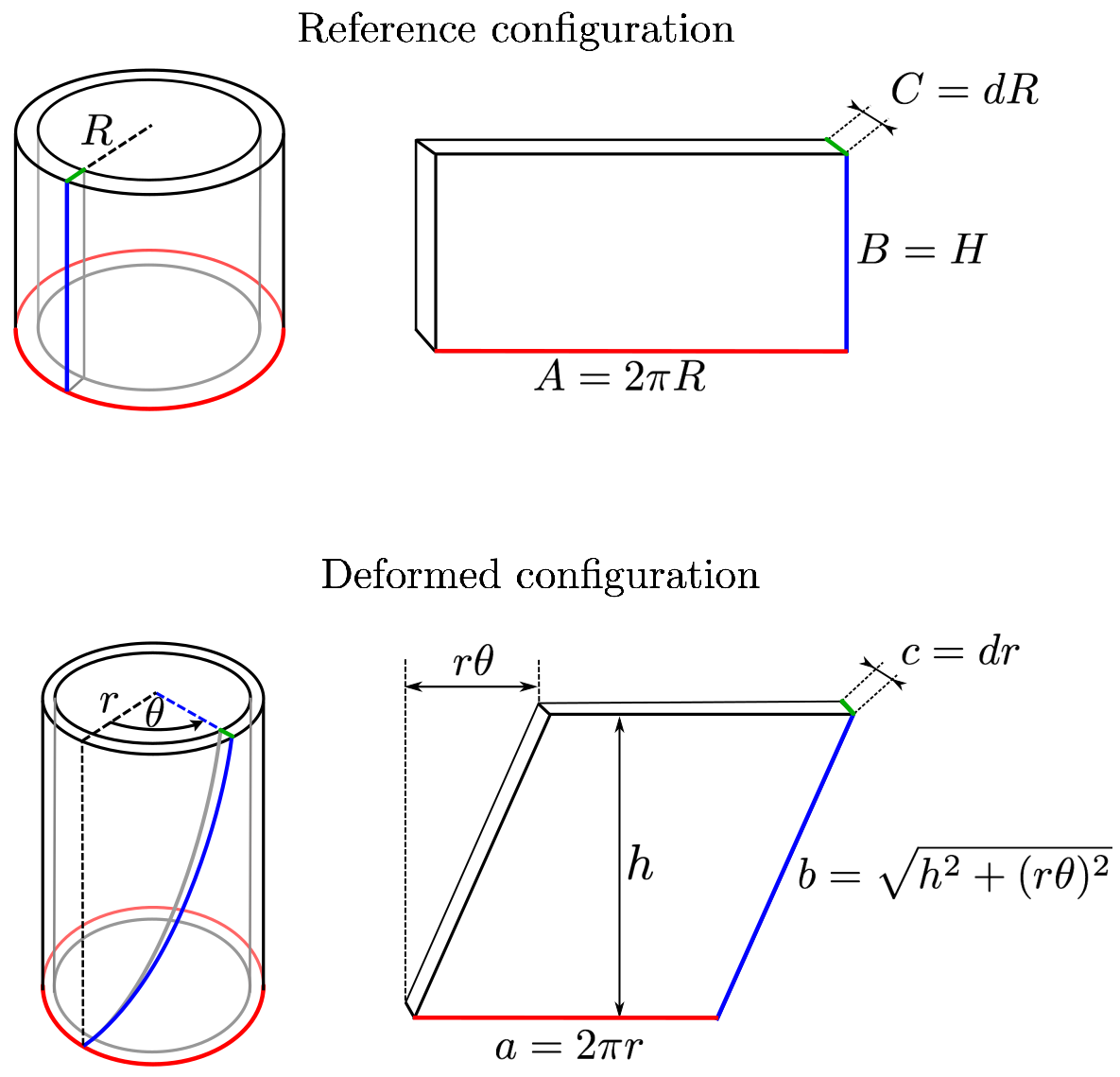}
\caption{\label{scheme}}
\end{centering}
\end{figure}

Consider an undeformed right parallelepiped with edges of lengths $A$, $B$, $C$, and assume it deforms into a parallelepiped with edges of lengths $a$, $b$, $c$. 
Based on a statistical thermodynamics analysis of cross-linked polymer chains, Treloar \cite{Treloar43} proposed  the following generalisation of the Hookean one-dimensional spring energy $U=\tfrac{1}{2}k(x-x_0)^2$ to model the behaviour of a soft solid undergoing such a deformation,
\beq\label{UU}
U = \frac{\mu}{2}\left(\frac{a^2}{A^2}+\frac{b^2}{B^2}+\frac{c^2}{C^2} - 3\right)V,
\eeq
where $V = ABC$ is the initial volume.
Here, $\mu >0$ is the  {\it shear modulus},  the analogue of the spring stiffness $k$. 
We shall use this energy to analyse the torsion of a soft solid.

We follow the context of Poynting's second paper on the twisting of cylinders \cite{Poynting1913}, which focused on ``India-rubber cords". 
Rubber, and a host of other solids, displays \textit{incompressibility}, which means that its volume remains unchanged when deformed. 
It turns out that this property simplifies  the mathematical analysis greatly. 

We call $H$ the height of our cylinder and  $R_{e}$ its radius when it is not subjected to forces, in its {\it initial configuration}. 
Once twisted by an angle $\theta$, it may change its height to $h$ and its radius to $r_{e}$, in its \textit{final configuration}. 

To compute the elastic energy stored in the cylinder during its deformation, we first look at a  cylindrical shell  of initial infinitesimal thickness $dR$, deformed into a final shell of thickness $dr$, see Fig.\ref{scheme}. 
Because rubber is incompressible, $dV$, the volume of the infinitesimally thin tube is conserved during the deformation, so that $dV = (2\pi R)H dR = (2\pi r) h dr$. 
By integration of this identity we thus obtain
\beq
H\int_0^{R}\tilde{R}\,d\tilde{R} = h\int_0^{r}\tilde{r}\,d\tilde{r}\quad\Rightarrow\quad
r=\sqrt{\frac{H}{h}}R. 
\label{r-R}
\eeq
When evaluated at the external radii $R_e$, $r_e$, the expression above tell us that, as expected, if the cylinder elongates during torsion, it must become thinner, and vice versa if it shortens. 

To compute the elastic energy of the cylinder in the torsion experiment we {\it unfold} the infinitesimally thin initial and final tubes into parallelepipeds, as done in Fig.\ref{scheme}, to establish the identifications 
\beq
\left\{
\begin{array}{lll}
A=2\pi R\\
B=H\\
C=dR
\end{array}
\right.
\qquad
\left\{
\begin{array}{lll}
a=2\pi r\\
b=\sqrt{h^2+(r\theta)^2}{}\\
c=dr
\end{array}
\right.
\eeq
Using \eqref{UU} and  \eqref{r-R}, we then find that the elastic energy of the infinitesimal volume $dV = 2\pi HRdR$ is
\beq
dU={\mu} \pi H
\left(2\frac{H}{h} + \frac{h^2}{H^2} + \frac{R^2\theta^2}{Hh} - 3\right) R dR. 
\eeq
To obtain the total elastic energy in the cylinder, we can integrate this quantity so that finally,
\beq
U=\int_0^{R_e}dU = \mu\pi H\left[\left(2\frac{H}{h} + \frac{h^2}{H^2} - 3 \right)\frac{R_e^2}{2} + \frac{\theta^2}{Hh}\frac{R_e^4}{4}\right]. 
\eeq

We can now deal with different scenarios, depending on the force and/or amount of twist that we impose on the cylinder. 

The first situation is the one described in Fig.\ref{cylinders}, where ideally only a twist is prescribed. 
In practice we can imagine that a bar with negligible mass is glued at the bottom of the cylinder and rotated  to twist it. 
Then the energy is $E = E(h) = U(h)$, and the energy minimiser  is found by solving
\beq
E'(h) = \mu\pi R_e^2\left[-\frac{H^2}{h^2} + \frac{h}{H}  - \frac{R_e^2\theta^2}{4h^2}\right]=0, 
\eeq
and  by imposing that
\beq
E''(h) = \mu\pi {R_e^2}\left[2\frac{H}{h^3} + \frac{1}{H}  + \frac{ R_e^2 \theta^2}{2h^3}\right] \geq 0.
\eeq
The former condition gives the equilibrium height $h^\star$ of the cylinder resulting from a twist of angle $\theta$ between its upper and lower faces, as 
\beq\label{heq}
h^\star = H \sqrt[3]{1+ \frac{R_e^2\theta^2}{4H^2}},
\eeq
and the latter, which is always satisfied, secures that $h^\star$ is indeed an energy minimiser. The expression \eqref{heq}  proves the result found by Poynting: \textit{a cylinder always elongates when subjected to torsion only}, because clearly, $h^\star > H$. 

For a thin and long cylinder ($R_e/H \ll 1$), the expression above can be expanded as
\beq\label{heqslender}
h^\star  \simeq H\left(1+\frac{R_e^2\theta^2}{12 H^2}\right). 
\eeq
The analysis of \eqref{heq} and \eqref{heqslender} shows that $h$ depends nonlinearly on $\theta$ both in stubby and in slender cylinders, confirming that the Poynting effect is a non-linear effect (no linear term in the expansion), and that the lengthening of the bar $h^\star-H$ is approximately proportional to the square of the diameter, as observed experimentally by Poynting \cite{Poynting1909, Poynting1913}. 

The second scenario is the one more likely to be encountered in the class: when a constant force $F$ is applied to stretch the rubber cord, by attaching a weight at its bottom end. Now  the work term is $W=F(h-H)$. In this case, energy minimisation yields an implicit equation for the height  at equilibrim
\beq
\left(\dfrac{h}{H}\right)^3 -\frac{F}{\mu \pi R_e^2}\left(\dfrac{h}{H}\right)^2 -  \dfrac{R_e^2\theta^2}{4H^2} -1= 0. 
\label{withF}
\eeq
Also in this case we find the Poynting effect, i.e. that $h^\star>H$, see Section \ref{Suggested problems}.

Another scenario takes place when a moment $M$ is imposed to create the twist. 
In that case, we can easily prove Coulomb's result that $M \thicksim \theta R_e^4/h$, see Section \ref{Suggested problems} for the details.


\section{Classroom experiment}


It is quite simple to create a classroom demonstration to provide a clear illustration of the Poynting effect for a rubber cylinder undergoing torsion (see linked 
\href{https://www.youtube.com/watch?v=ugD6PsDaLu4}{\color{blue}\underline{movie}}\color{black}.) 
\begin{figure}[h!]
\centering
\includegraphics[width=0.45\textwidth]{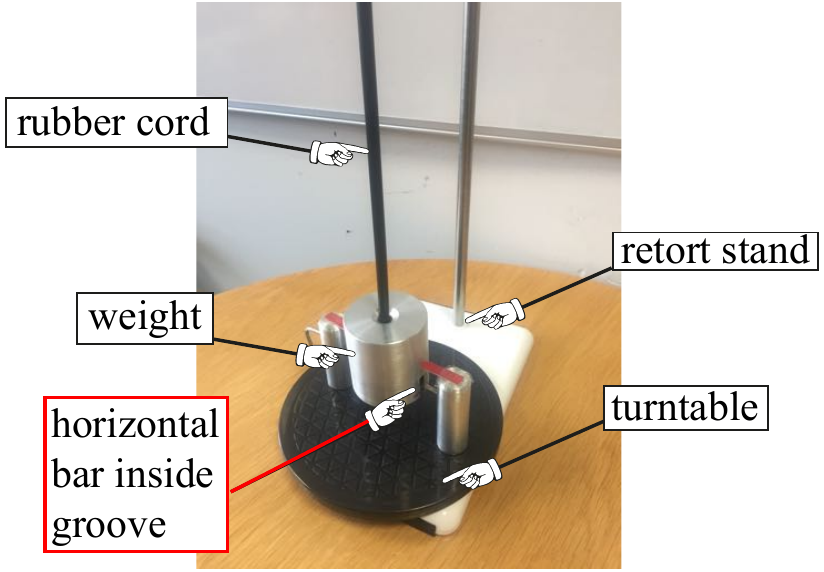}
\caption{A simple setup to demonstrate the Poynting effect}
\label{fig:setup}
\end{figure}

In our experiment, a 1 m long, solid cylinder of Nitrile 70 rubber (Camthorne Industrial Supplies) with 8 mm diameter is held vertically by its top end using a retort stand, and an aluminum metal weight with gradation etchings is clamped at its bottom.
The  weight has a groove into which a horizontal bar can lodge.
The horizontal bar is linked to a turntable, used to rotate the weight and thus twist the rubber cord, see Fig.\ref{fig:setup}.

\begin{figure}[h!]
\centering
\includegraphics[width=0.45\textwidth]{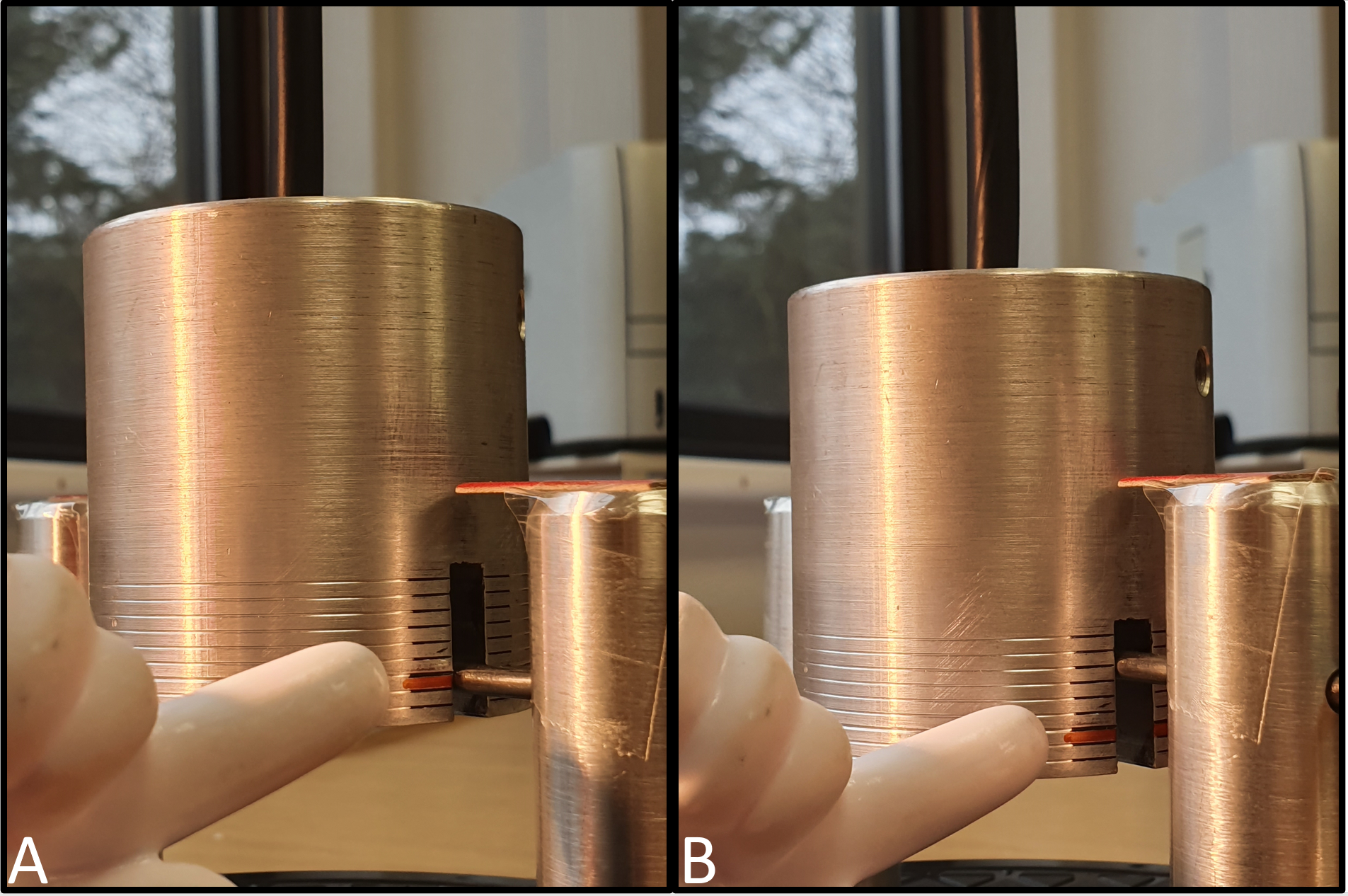}
\caption{(A) Initial altitude of the aluminium cylinder attached at the end of the rubber cord, as indicated by the alignment of the red mark  with the horizontal bar. (B) After six  full turns of the cord, the  cylinder has clearly gone down, showing that the cord elongated with the torsion. For more details, watch the linked
\href{https://www.youtube.com/watch?v=ugD6PsDaLu4}{movie}}
\label{fig:experiment}
\end{figure}

First, the gradation markings at the equilibrium position are highlighted to show the original altitude of the weight with respect to the horizontal bar.  
Then the turntable is rotated: after several rotations the position of the gradations is checked against the position of the horizontal bar, and it is clear that the cylindrical rubber cord has lengthened with respect to the original markings, see Fig.\ref{fig:experiment}.

In our experiments, the cylinder is rotated 6 times and the elongation of the cord is measured. 
As seen on Fig.\ref{fig:experiment}, the elongation is measured to be about 4.5 marks. 
The gradation marks are in steps of 2 mm which gives a length increase of approximately 9 mm. 
The aluminum weight weighs about $\sim$ $0.3$ kg,  which imposes a force $F$ on the cord. 
In the end we find that Eq.~\eqref{withF} predicts an elongation of $\sim 9$ mm, in good agreement with the observation (details are left as an exercise in Section \ref{Suggested problems}.)


\section{Further reading}


John Henry Poynting FRS was the professor of physics at the University of Birmingham  until his death in 1914. He is mostly remembered for the ``Poynting vector'', which gives the direction and magnitude of the electromagnetic flux. 
In fact, searching of his name in the pages of this journal yields hundreds of articles on the vector that bears his name, but none on the intriguing non-linear elastic effect which he discovered in 1909, and is the subject of this article.

Torsion is a core topic of elasticity. Eminent figures such as Cavendish, Coulomb,  Saint-Venant and Prandtl  researched it. 
Yet the question ``does a  cylinder shorten, keep its height, or elongate when twisted?'' remained unanswered until Poynting's experiments on metal wires and rubber cords. 
A fully analytical modelling of this effect appeared only in 1947, and the answer to the question still remains non-intuitive to most, including scientifically-minded people.

Nowadays, the Poynting effect is a classical topic of non-linear elasticity and is  taught in third-level education. Its proof, however, is rather technical, and is generally postponed to the end of advanced elasticity modules, because it relies on tensor algebra, see for example the classic books by Truesdell and Noll \cite{Truesdell} or Ogden \cite{Ogden}. 
To make the derivation more accessible, here we took an alternative route based on energy minimization, that builds on basic knowledge of advanced high school or first-year university single variable calculus and physics. A slight twist (so to speak) in the problem that requires knowledge of multivariable calculus and algebra is proposed as an exercise in Section \ref{Suggested problems}, but it still remains fully accessible to early third-level students. 

In practice, the Poynting effect provides a simple way of determining experimentally the elastic shear modulus $\mu$ of soft matter, see the formulas derived in the exercises of Section \ref{Suggested problems}. 
It is a useful alternative to other protocols based for example on tensile tests \cite{Allen}, inflation of a rubber torus \cite{Bruce92}, oscillations \cite{Papadakis}, etc.
With the recent development of very precise rheometers, it can even be exploited to measure the elastic constants of extremely soft solids, such as brain matter \cite{Balbi}.


\section{Suggested problems}
\label{Suggested problems}


\begin{enumerate}
\item
Equation \eqref{withF} can be written 
\[
f(x) \equiv x^3-ax^2-b-1=0,
\] where 
\[
x= \dfrac{h}{H}, \qquad a = \dfrac{F}{\mu \pi R_e^2}, \qquad b = \dfrac{R_e^2\theta^2}{4H^2}.
\]
Study the variations of $f$ to show that it has only one real root $x^\star$, which is such that $x^\star>1$,  hence showing the Poynting effect.

\item
If the pre-tensioning force is small compared to the stiffness of the rubber, i.e. $a \ll 1$, show that $x^* \simeq \sqrt[3]{b+1} + a/3$, which is to say that
\[
h^\star \simeq H \left(\sqrt[3]{1+ \frac{R_e^2\theta^2}{4H^2}} + \dfrac{F}{3\mu \pi R_e^2}\right).
\]

\item When an applied  torque $M$ is the control variable instead of the twist $\theta$, then $\theta$ becomes an independent variable which needs to be calculated. 
In this case there is a contribution $M\theta$ in the work term and the total potential energy becomes a function of two variables $U=U(h,\theta)$. 
Now the equilibrium conditions \eqref{cond} can be recast as
\[
\frac{\partial U}{\partial h} = 0, \qquad \frac{\partial U}{\partial \theta} = 0,
\]
together with the requirement that the so-called \textit{Hessian matrix} 
\[
\begin{bmatrix}
\frac{\partial^2U}{\partial h^2} & \frac{\partial^2U}{\partial h\partial\theta}\\[4pt]
\frac{\partial^2U}{\partial h\partial\theta} & \frac{\partial^2U}{\partial\theta^2}
\end{bmatrix}
\quad \text{is positive semi-definite}.
\]
Show that the elongation is now found by solving 
\[
\left(\dfrac{h}{H}\right)^3 -\left(\frac{F}{\mu \pi R_e^2} +   
\dfrac{M^2}{\mu^2\pi^2 R_e^6}\right)\left(\dfrac{h}{H}\right)^2 -1= 0,
\label{withMt}
\]
and that the twist is
\[
\theta=\dfrac{2 h M}{\mu \pi R_e^4}. 
\]
Similar to the first exercise, show that $h^\star>H$ here too.
Show, furthermore, that the Hessian matrix corresponding to this solution is indeed positive semi-definite.

\item

Check the correctness of our estimate of the elongation of the twisted cord by solving Eq.~\eqref{withF} with the following data: the cord has a stress-free length of 1 m, a diameter of 8 mm, and a shear modulus $\mu=400$ psi \cite{KC}; the aluminum cylinder has a mass of $300$ g; and we twisted the cord 6 full turns. Be careful with dimensions!

\end{enumerate}





\begin{thebibliography}{100}


\bibitem{Allen}
Allen B., 
From the simple to the surprisingly complex--An incremental study of elasticity. The Physics Teacher, 57, 570-571 (2019).

\bibitem{Balbi}
Balbi V., Trotta A., Destrade M., and Annaidh A. N.  
Poynting effect of brain matter in torsion. 
Soft matter, 15, 5147-5153 (2019).

\bibitem{Bruce92}
Bruce I., 
The elastic constant of a rubber tube. American Journal of Physics, 60, 157-160 (1992).

\bibitem{Coulomb}
Coulomb C-.A., 
Recherches th\'eoriques et exp\'erimentales sur la force de torsion et sur l'\'elasticit\'e des fils de metal,
Histoire de l'Acad\'emie Royale des Sciences, 229-269 (1784).


\bibitem{Britannica}
 \textit{Encyclopedia Britannica} Volume 12 (11th ed.), Chisholm H (ed.),  University Press, Cambridge (1911). 

\bibitem{KC}
KC Seals.
Material Properties: General Purpose NBR (N70, Nitrile 70 Duro, Buna), 
\href{https://kcseals.ca/uploads/images/pdfs/NBR_70.pdf}{https://kcseals.ca/uploads/images/pdfs/NBR\_70.pdf}

\bibitem{Ogden}
Ogden R.W., 
\textit{Non-Linear Elastic Deformations}, Ellis Horwood, Chichester (1984). Reprinted by Dover (1997).

\bibitem{Papadakis}
Papadakis E. P., 
Undergraduate experiment on elasticity of rubber bands. 
American Journal of Physics, 31, 938-939 (1963).

\bibitem{Poynting1909} 
Poynting J. H., 
On pressure perpendicular to the shear-planes in finite pure shears, and on the lengthening of loaded wires when twisted, Proceedings of the Royal Society A 82, 546-559 (1909). 

\bibitem{Poynting1913} 
Poynting J. H., 
The changes in length and volume of an Indian-rubber cord when twisted, India-Rubber Journal, October 4, p.6 (1913). 

\bibitem{Rivlin47}
Rivlin R. S., 
Torsion of a rubber cylinder. Journal of Applied Physics 18, 444-449 (1947).

\bibitem{Treloar43}
Treloar L. R.G., 
The elasticity of a network of long-chain molecules. 
Transactions of the Faraday Society, 39, 36-41 (1943).

\bibitem{Truesdell}
Truesdell C. and Noll W.,  
\textit{The Non-Linear Field Theories of Mechanics}. Springer, Berlin (2004).

\end{thebibliography}
\end{document}